\documentclass{article}
\usepackage{spconf,amsmath,graphicx, amssymb, cite, bm, caption, booktabs, lipsum, subcaption, multirow}
\usepackage[breaklinks=true]{hyperref}

\newcommand\linesubsec[1]{\vspace{0.8mm}\noindent\textbf{#1 --- }}

\title{SYNCFUSION: MULTIMODAL ONSET-SYNCHRONIZED\\VIDEO-TO-AUDIO FOLEY SYNTHESIS}
\name{
\textit{Marco Comunità$^{1*}$ \qquad Riccardo F. Gramaccioni$^{2*}$ \qquad Emilian Postolache$^{2}$}
\thanks{\hspace{-6pt} $^*$ equal contribution}\vspace{0.2cm}\\
\textit{Emanuele Rodolà$^2$ \qquad Danilo Comminiello$^2$ \qquad Joshua D. Reiss$^1$}\\
}
\address{
$^1$Centre for Digital Music, Queen Mary University of London, UK\\
$^2$Sapienza University of Rome, Italy
}

\begin{document}
\ninept
\maketitle
\begin{abstract}
Sound design involves creatively selecting, recording, and editing sound effects for various media like cinema, video games, and virtual/augmented reality. One of the most time-consuming steps when designing sound is synchronizing audio with video. In some cases, environmental recordings from video shoots are available, which can aid in the process. However, in video games and animations, no reference audio exists, requiring manual annotation of event timings from the video. 
We propose a system to extract repetitive actions onsets from a video, which are then used - in conjunction with audio or textual embeddings - to condition a diffusion model trained to generate a new synchronized sound effects audio track. In this way, we leave complete creative control to the sound designer while removing the burden of synchronization with video. Furthermore, editing the onset track or changing the conditioning embedding requires much less effort than editing the audio track itself, simplifying the sonification process. We provide sound examples, source code, and pretrained models to faciliate reproducibility\footnote{\href{https://mcomunita.github.io/diffusion-sfx\_page}{https://mcomunita.github.io/diffusion-sfx\_page}}.

\end{abstract}

\begin{keywords}
Sound effects synthesis, foley, diffusion models, audio-video synchronization, multimodal audio synthesis.
\end{keywords}

\vspace{-4pt}
\section{Introduction}\label{sec:intro}
\vspace{-4pt}


Sound plays an essential part in the narration of any audiovisual work. Consequently, it should come as no surprise that the role of the Foley artist - who creates sound effects for films, video games, commercials, etc. - is crucial to achieving top-quality productions.
This task presents considerable difficulties, as it is necessary to create an audio track that corresponds perfectly both in time and content to the video to be soundtracked.
In contexts like video games and animated movies, sound designers often receive silent videos, requiring them to create soundtracks entirely from scratch without timing guidance.
In other cases, like cinematographic filming, a raw audio track may accompany the video, but Foley artists can only rely on it for timing, while sounds have to be recreated from the ground up; often adopting totally different materials to those present in the video in order to seek a hyper-realism that can benefit the narration.
In any case, the essence of the work lies in sourcing and creating high-quality sounds, leading them to build distinctive sound libraries used across their productions.

\begin{figure}[t]
\centerline{\includegraphics[width=3.3in]{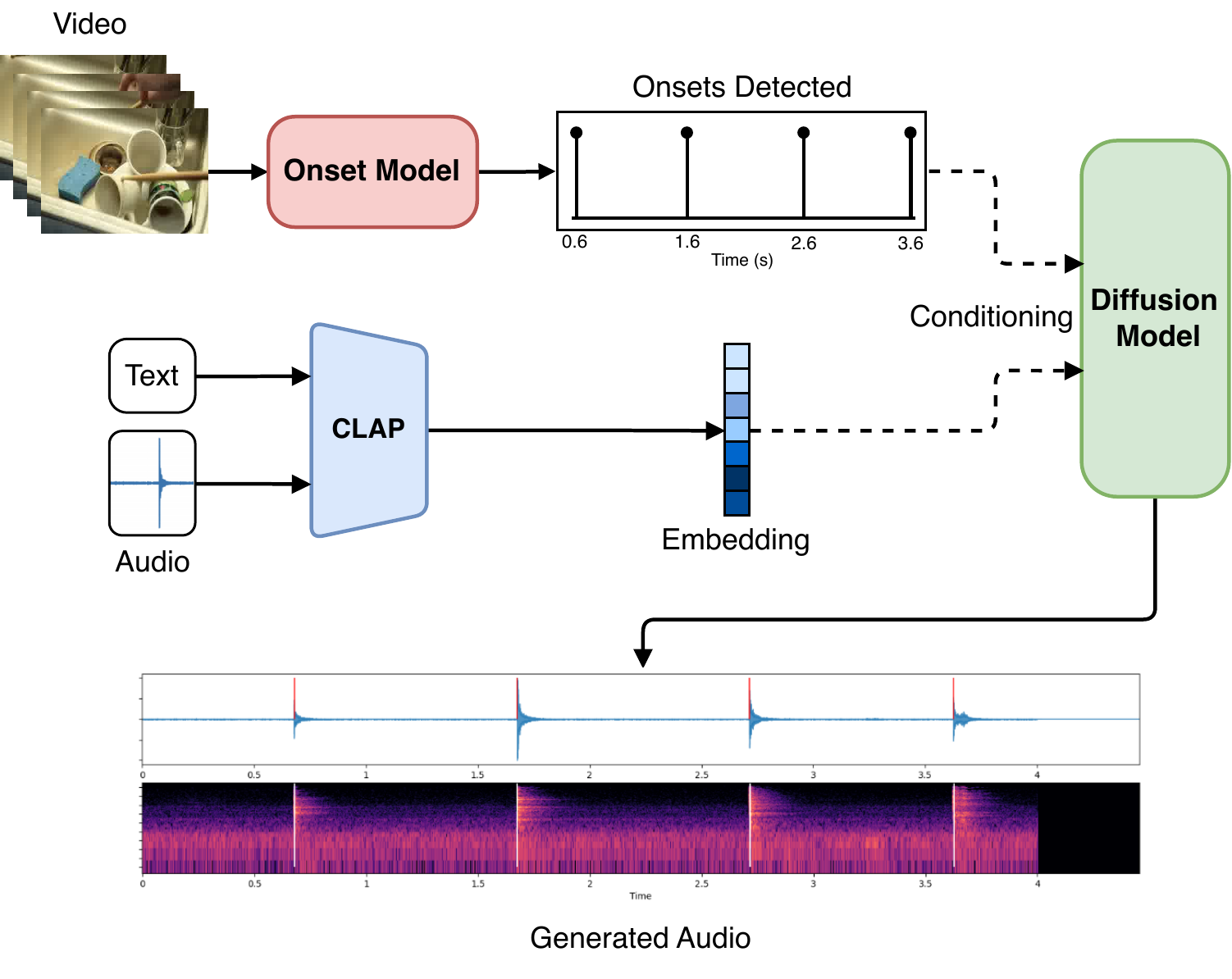}}
\vspace{-0.3cm}
\caption{\textit{
The overall architecture consists of two distinct parts: 1) an Onset Model that, 
given a silent video extracts the onsets for the actions in that video; and 2) a Diffusion Model which, conditioned on the onsets track and a CLAP embedding, generates a synchronized audio that can be used as soundtrack for the input video}}
\label{blockdiag}
\vspace{-0.3cm}
\end{figure}

While the creative aspect is highly stimulating, sound designers dedicate the majority of their time to meticulous, repetitive tasks, such as precise synchronization of sound events with video moments, crucial for maintaining viewer immersion.

Many works have tried to automate audio-video synchronization: \cite{Iashin2022SparseIS} propose a transformer-based architecture that allows the analysis of long video sequences; similar solutions have been adopted in \cite{Chen2021AudioVisualSI}, while in \cite{Kadandale2022VocaLiSTAA} the analysis is extended to lip synchronization for singing voices. 
In \cite{Dassani2019AutomatedCO} the aim is to generate a musical soundtrack that is synchronized with the movie pictures. 

\begin{figure*}[t]
    \vspace{-1cm}
    \centering
    \includegraphics[width=0.85\textwidth, height=6cm]{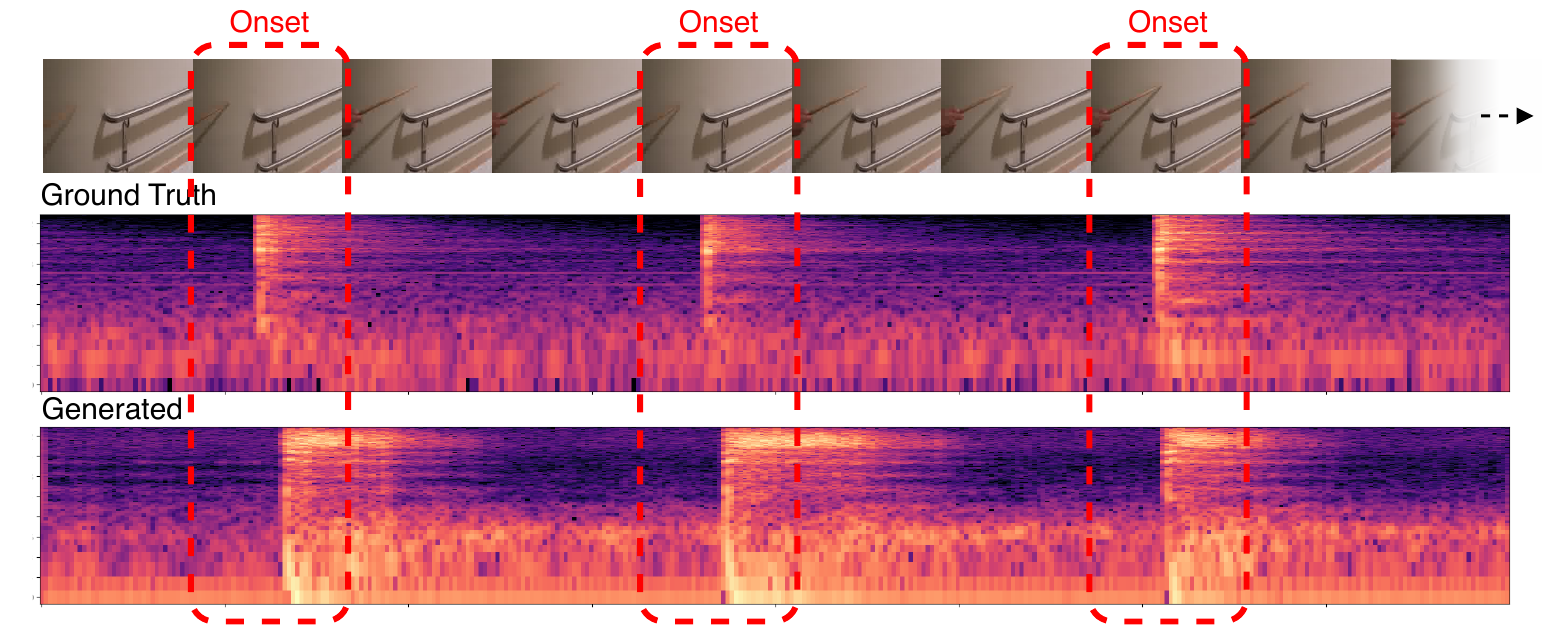}
    \caption{\textit{Example showing ground truth audio and video, detected onsets and generated audio}.}
    \label{fig:spec}
    \vspace{-0.2cm}
\end{figure*}

In sound design, the target is to create an ambient soundtrack that perfectly describes the scene in terms of general mood of the audiovisual work, while following the scenes' transitions as well as temporal and spatial localization.
Spatio-temporal event localization is a key computer vision task, that many research studies have tried to solve through the use of deep learning techniques. 
However, works often focused on detecting and counting the number of repetitions of a particular action in a video \cite{Zhang2021RepetitiveAC, 
Yao2023PoseRACPS}. 


In \cite{Dwibedi2020CountingOT} the repetition count is class agnostic, 
which is a case of major interest for the analysis of movie sequences, video games, and so forth. 
However, these works do not provide a precise timing of the actions, which is crucial information for soundtracking.


In recent years, video-to-audio tasks have started gathering wider attention and some works analyzed the generation of audio tracks that are temporally and thematically aligned to a given video sequence \cite{
Ghose2021FoleyGANVG, Su2023PhysicsDrivenDM}. 
Contrastive learning is providing remarkable results in domain translation from video to audio  for solving this challenging problem \cite{Luo2023DiffFoleySV, Wang2023V2AMapperAL}.
The authors of \cite{Du2023ConditionalGO} proposed a model to generate a soundtrack for a silent input video, given a reference audio-video pair that specifies what the video should sound like.

Compared to previous work, conditioning the audio generation model with onsets of the actions to be soundtracked can provide Foley artists with greater creative control, allowing them to bypass the mechanical task of manually annotating each repetition of the relevant action and focus exclusively on the quality of the sounds to be produced. Furthermore, modifying an onsets track is very simple and can be of great help when editing.
Therefore, we decided to base our work on a model that is able to perform onset detection of the actions present in a silent input video

Next, we developed a diffusion model that, conditioned on an embedding of the sound representing how the actions in the video should sound like and an onsets track depicting when those actions occur over time, generates an audio track that is in time and content aligned with the input video.
This choice is motivated by the extraordinary results that diffusion models have recently achieved in audio generation \cite{Mariani2023MultiSourceDM, Pascual2022FullbandGA 
}. 
A block diagram of the proposed system is shown in Figure \ref{blockdiag}.



\vspace{-4pt}
\section{Method}\label{sec:method}
\vspace{-4pt}


In this section we detail the different components of the proposed model. The video onset network to detect the occurrences of the relevant actions in the silent input video. The audio representation network to obtain an embedding of the desired target sound from audio or textual source. Finally, the diffusion model which, conditioned on onset track and sound embedding, generates the onset-synchronized audio track.




\vspace{-4pt}
\subsection{Video Onset Detection}
\vspace{-4pt}

Inspired by the work done in \cite{Du2023ConditionalGO}, we selected a ResNet(2+1)D-18
as the video onset detection network.
Following their implementation, we removed all temporal striding so that the last convolutional layer would have the same temporal sampling rate as the input video and therefore preserve more detailed temporal information.

At the final stage, after pooling, a fully connected layer outputs a label of the same length in frames as the input video. Each element of this label is a prediction representing the presence or absence of a given action at the specific frame. Consequently the resulting label will be a binary mask in which the value $1$ for the \textit{i}-th element represents the presence of an onset for frame \textit{i}, while the value $0$ indicates the fact that no action has been detected for that frame. Therefore, given a silent input video $V \in \mathbb{R}^{C \times H \times W \times T}$, where $C$ is the number of input channels, $H \times W$ is the dimension in height and width of each frame and $T$ is the total duration of the video expressed in frames, the video onset model outputs an onset label $o \in \mathbb{R}^T$ where each element $o_i$ is defined as:

\begin{equation*}
o_i =
\left\{ 
\begin{array}{rl}
1, & \mbox{if there is an action in the \textit{i}-th frame}\\
0, & \mbox{if there is no action in the \textit{i}-th frame}
\end{array}
\right.
\end{equation*}




\vspace{-4pt}
\subsection{Audio Representation}
\vspace{-4pt}
In recent years there has been a substantial effort to develop general-purpose audio representations that generalize well to a variety of downstream tasks \cite{
niizumi2021byol}, with contrastive learning becoming a widely adopted training regime \cite{saeed2021contrastive}, especially in the case of multimodal approaches. A successful example of multimodal representation learning is CLAP \cite{elizalde2023clap}, where embeddings for the audio and text modalities are aligned in the latent space. 
We leverage such alignment conditioning our synthesis model on audio embeddings only at training time, allowing textual queries as a secondary conditioning modality at inference time.


\vspace{-4pt}
\subsection{Sound Effects Synthesis}
\vspace{-4pt}

We generate a time-domain sound effect sequence $\mathbf{x}(0)$ by employing a variance-preserving continuous-time diffusion model $S_\theta$ \cite{song2021scorebased, salimans2022progressive}, capturing the gradient of the noisy log-distribution:
\begin{equation*}
\label{eq:diffusion}
 \nabla_{\mathbf{x}(t)} \log p(\mathbf{x}(t)) \approx S_\theta(\mathbf{x}(t), \sigma(t)),
\end{equation*}
where $p(\mathbf{x}(t)) = \int_{\mathbf{x}(0)}p(\mathbf{x}(t) \mid \mathbf{x}(0))p(\mathbf{x}(0))$, with:
\begin{equation*} 
\label{eq:perturbation_kernel}p(\mathbf{x}(t) \mid  \mathbf{x}(0)) = \mathcal{N}(\mathbf{x}(t) \mid \alpha(t)\mathbf{x}(0), \sigma^2(t) \mathbf{I})
\end{equation*} a Gaussian perturbation kernel. Following \cite{salimans2022progressive}, we use a noise schedule $\sigma(t)\in [0, 1]$ and $\alpha(t) = \cos(0.5\pi \sigma(t))$. $S_\theta$ is trained by minimizing the following loss:
\begin{equation*}
\mathbb{E}_{\sigma(t) \sim [0,1], \mathbf{x}(t)}\left[ \Vert S_\theta(\mathbf{x}(t), \sigma(t)) - \mathbf{v}(t)\Vert^2_2 \right]\,,
\end{equation*}
where $\mathbf{x}(t) = \alpha(t)\mathbf{x}(0) + \beta(t)\bm{\epsilon}$, $\mathbf{v}(t) = \alpha(t)\bm{\epsilon} - \beta(t)\mathbf{x}(0)\,,$ with $\beta(t) = \sin(0.5\pi\sigma(t))$ and $\bm{\epsilon}$ white noise. 
For sampling, we use a standard DDIM integrator \cite{song2021denoising}. The architecture of $S_\theta$ is a UNet that follows the design of Moûsai \cite{schneider2023mo}. The encoder/decoder structure of the residual UNet has 8 layers 
and a total downsampling/upsampling factor of 1024. 
The innermost 4 layers perform self-attention with 8 attention heads and 64 features. 
To condition with CLAP embeddings we use cross-attention and train with classifier-free guidance \cite{ho2021classifierfree}. To condition the diffusion model with the onsets, we feed them to a convolutional encoder that follows the structure of the UNet encoder 
and inject the channels at the corresponding layer inside the UNet.


\vspace{-4pt}
\section{Experimental Design}\label{sec:exp}
\vspace{-4pt}

\subsection{Dataset}
To train and test our models we adopt the widely-used 
Greatest Hits dataset \cite{Owens2015VisuallyIS}.
This dataset includes videos of humans using a drumstick to hit or rub objects or surfaces. The choice of a drumstick as the striking object is useful, as it minimally occludes each frame, enabling the video onset detection network to better comprehend the scene.
Each video in the dataset captures the drumstick action, and the audio is recorded with a shotgun microphone attached to the camera, followed by denoising. 
The dataset contains onset annotations for each video, along with action and material labels for most events. This comprehensive dataset is fundamental for our model since it is the only one of sufficient size and quality, providing the audio-visual information our model relies on.

Altogether, the dataset consists of 977 videos captured both outdoor and indoor. Indoor scenes contain a variety of hard and soft materials, such as metal, plastic, cloth, while the outdoor scenes contain materials that scatter and deform, such as grass, leaves and water.
On average, each video contains 48 actions, divided between hitting and scratching.
This ensures that each extracted chunk of video, lasting either 2s (for the onset model) or 6s (for the synthesis model), contains a sufficient number of hits.

We divided the dataset into 683 videos for the training set, 98 for the validation set and 196 for the test set (70/10/20\%). 

\vspace{-4pt}
\subsection{Experiments}
\vspace{-4pt}
We split the problem of video-to-audio synthesis
into a video analysis stage and a sound synthesis stage. In order to establish which of these stages is the limiting factor on the overall performance, we organize training and evaluation in three main parts for video onset detection stage, sound effects synthesis stage, and complete system.
Evaluation of the complete system is conducted on pre-trained models, i.e., we do not attempt end-to-end training of both models.

As objective metrics - similar to previous work \cite{Du2023ConditionalGO} - we measure the accuracy on the number of detected/synthesized onsets and the average precision score\footnote{\href{https://scikit-learn.org/}{https://scikit-learn.org/}}, which measures the synchronization between models' outputs and ground truth.
To further evaluate how well the synthesized audio approximates the training data we use the Fréchet Audio Distance (FAD) \cite{kilgour2018fr} 
which, correlating with human judgment, is also a measure of perceived quality of individual sounds.


\linesubsec{Video Onset Detection} to assess the performance of the onset detection model we rely on ground truth annotations. Since the dataset includes - on average - an event every 1.5~s, we train the model on 2s long video chunks. Furthermore, to train more efficiently, we downsample the videos to a 15fps frame rate. To construct the input, we extract the single frames as image files, and feed the network groups of 30 consecutive frames. Overall, each batch has size $[B, T, C, H, W]$, with $B$ batch size, $T$ frames, $C$ color channels, $H$ and $W$ height and width of each frame.


We repeat the same experiment twice comparing the performance when using augmentations \cite{Perez2017TheEO} on the input frames. Without augmentations the frames are simply resized to a 112-by-112 dimension to match the model requirements, and are channel normalized with mean and standard deviation computed across the dataset for each color channel. When using augmentations, we first resize to 128-by-128 and apply random crop to the final size; we also apply color jitter before normalization. Augmentations are only applied to training and validation sets.

We train each model for 100 epochs on a binary cross-entropy loss with batch size of 16, using AdamW optimizer with weight decay of $1\cdot10^{-3}$ and a learning rate of $1\cdot10^{-4}$. 
To generate the output binary labels we use a sigmoid on the network output and apply a threshold of 0.5. We compute accuracy and average precision score at 15fps.

\linesubsec{Sound Effects Synthesis} to train the diffusion model we use audio batches - extracted from the videos - of shape $[B, C, L]$, with $B$ batch size, $C$ audio channels, and $L$ length in samples. The model trains on windows of $2^{18}$ samples at 48kHz ($\sim$6s). 
Ground truth onset annotations are used to build binary tensors at audio rate for conditioning.
We also 
zero out the audio track before the first onset to remove possible tails from previous events. Finally, a conditioning segment is extracted
by randomly choosing an onset 
and slicing until the following one.
Such slice is embedded with CLAP and the result is fed to the UNet via cross-attention. For classifier-free guidance, during training, we use a constant embedding with 0.1 probability, and for inference, we use an embedding scale of 2.

The model is trained with the AdamW optimizer with a batch size of 2 for 1000 epochs with weight decay $1\cdot 10^{-3}$ and a learning rate of $1\cdot 10^{-4}$.


To evaluate our model, we create 2-second clips using the initial ground truth onsets from the test set videos. We exclude tracks with zero onsets, leaving us with 160 segments. To account for any potential initial onset misses in the manual annotations of the Greatest Hits dataset, we reset the beginning of both the generated and ground truth tracks until the first annotated event.

Differently from the onset model, in this case we compute the objective metrics at audio sample rate, using a confidence interval of 50ms.
Furthermore, we compute the FAD for both audio and text modalities using the labels available in the dataset as text queries. 


\linesubsec{Complete System} performance of the complete system is measured by first generating the binary labels with the pre-trained onset networks and converting them into onset tracks at audio sample rate to condition the synthesis model. Onsets are then extracted from the ground truth and synthesized audio using \textit{librosa} \footnote{\href{https://librosa.org/doc/0.10.0/index.html}{https://librosa.org/doc/0.10.0/index.html}}, and a tolerance of $\pm$50ms is applied to compute the onset synchronization precision. Although we have ground truth annotations, we adopt this approach for a fair comparison with the baseline described below, which does not use annotations. We generate 160 chunks like in the previous scenario. 
Even if \textit{librosa}'s onset detection tool has mainly been used for the peak detection of musical audio segments, here the direct application of this feature is justified by the fact that ground truth and generated audio tracks contain minimal if any background noise, allowing for a correct onsets extraction without further processing steps.

An example of the proposed model's output is shown in Fig. \ref{fig:spec}.

\linesubsec{Baseline} we compare our approach with a recent work \cite{Du2023ConditionalGO} where a model is proposed to sonify a silent video using a conditioning audio-visual pair. 
CondFoleyGen uses SpecVQGAN \cite{SpecVQGAN_Iashin_2021} to learn a codebook from the training data spectrograms. 
During training, the code for the conditioning audio is passed as input - along with the embeddings for the conditioning and silent videos - to a transformer, which autoregressively predicts the codes that should represent the target sound. Then, a MelGAN vocoder \cite{Kumar2019MelGANGA} produces the waveform generated from the spectrogram reconstructed by the codebook decoder.
Finally, an audio-visual synchronization model is used to re-rank many generated soundtracks and choose the one with the best temporal alignment.

Since no pre-trained models were available at the moment of writing, we re-trained the model using the code provided in the official repository\footnote{\href{https://github.com/XYPB/CondFoleyGen/tree/main}{https://github.com/XYPB/CondFoleyGen/tree/main}} and following details from the paper.
Accordingly, we trained the SpecVQGAN codebook for 400 epochs and the transformer for 40 epochs in order to make a fair comparison with our model.





\vspace{-4pt}
\section{Results}\label{sec:results}
\vspace{-4pt}

\linesubsec{Video Onset Detection} 
Table \ref{tab:onset} shows the results for the onset detection model in the two cases. 
As a fairly simple architecture, not specifically designed for the task at hand, the overall results are satisfying, with augmentation improving
both accuracy and average precision. In either cases, the model seems to be well suited to condition our synthesis model. We noticed that the network tends to overestimate the number of detected events, which becomes the main limitation in terms of reliability for the overall system.


\begin{table}[t]
    \centering
    \caption{Onset detection model evaluation}
    \begin{tabular}{l c c c c c} \toprule
                                & Params.   & Loss      & \# Onset & Onset Sync.  \\ 
                                &           & BCE       & Acc. (\%) $\uparrow$ & AP (\%) $\uparrow$ \\ 
        \midrule
        w/out augm.             & 31M     & 3.79          & 43.07    & 87.71 \\
        w/ augm.                & 31M     & \textbf{2.64} & \textbf{49.39}    & \textbf{88.83} \\
        \bottomrule 
    \end{tabular}
    \vspace{-0.0cm}
    
    \label{tab:onset} \vspace{0.3cm}
\end{table}

\linesubsec{Sound Effects Synthesis} 
Table \ref{tab:diffusion} reports the evaluation for the diffusion model. Accuracy and average precision are computed between conditioning onset tracks and generated output, with a tolerance of $\pm$50ms. FAD is computed between conditioning and synthesized audio.
Metrics are measured for both, audio and text modalities, to measure the impact of training with audio modality only.

The model learns very well to synchronize generated audio with the onsets. This 
supports the idea of separating video analysis and audio synthesis tasks; knowing that, upon development of better video understanding models, a more precise conditioning could be given for synthesis in terms of both timing and target sound.

The FAD for the two modalities is very similar, highlighting the alignment in the embedding space.


\linesubsec{Complete System}
results for the complete model are shown in Table \ref{tab:complete}. Comparing with Table \ref{tab:diffusion}, we notice how the onset detection model is the limiting factor in terms of accuracy and synchronization. 
In this case, augmentations have a negative impact, but the objective metrics remain in line with Table \ref{tab:onset}.
With respect to the baseline, the strong conditioning induced by the onsets track results in higher performance with almost half the number of parameters.

The increase in onset accuracy with respect to the onset model alone might be explained by the tendency of the onset model to overestimate the number of onsets. When detected from the audio, nearby false positives in the conditioning onset track might be ``obscured'' by preceding ones, leading to a few percentage point improvement. 

The proposed system is also surpassing the baseline - although by a limited margin - in terms of audio quality when measured with the FAD. Again, with respect to the diffusion model alone, we observe also a degradation in the FAD. This can be explained by the fact that in the complete system, we do not zero out the generated and test tracks like with the standalone diffusion model, resulting in a lower correlation between ground truth and generated tracks.

\begin{table}[t]
    \centering
    \caption{Synthesis diffusion model evaluation}
    \begin{tabular}{l c c c c c} \toprule
        Modality            & Params.   & \# Onset              & Onset Sync.   & FAD $\downarrow$ \\ 
                            &           & Acc. (\%) $\uparrow$  & AP (\%) $\uparrow$       & \\ 
        \midrule
        Audio               & 215M      & \textbf{89.38}            & \textbf{98.75}             & \textbf{1.48} \\
        Text                & 215M      & 84.38             & 98.12             & 1.68 \\ 
        \bottomrule 
    \end{tabular}
    \vspace{0.2cm}
    \label{tab:diffusion}
\end{table}

\begin{table}[t]
    \centering
    \caption{Complete model (SyncFusion) evaluation}
     \resizebox{1.\linewidth}{!}{%
    \begin{tabular}{l c c c c c} \toprule
        Model               & Params.   & \# Onset      & Onset Sync.   & FAD $\downarrow$\\ 
                            &           & Acc. (\%) $\uparrow$     & AP (\%) $\uparrow$       & \\ 
        \midrule
        Ours (w/out augm.)  & \textbf{246M}     & \textbf{56.87}             & \textbf{84.37}             & 5.50 \\
        Ours (w/ augm.)     & 246M     & 49.38             & 79.11             & \textbf{5.38} \\
        CondFoleyGen        & 408M      & 23.94             & 62.44             &  6.10 \\ 
        \bottomrule 
    \end{tabular}
    }%
    \vspace{-0.0cm}
    \label{tab:complete}
\end{table}

\vspace{-4pt}
\section{Discussion}\label{sec:Discussion}
\vspace{-4pt}
Video-to-audio tasks are garnering researchers' attention, thanks to the rapid advancement of both vision and audio generation architectures.
Even if sound effects and environmental sounds are crucial for the task, these applications are lagging behind speech and music synthesis, especially in professional sound design, where high-quality audio libraries and annotations are vital.
In fact, no datasets specific for Foley generation tasks are available, with Greatest Hits \cite{Owens2015VisuallyIS} being the only exception. 
This choice allowed us to compare with the selected baseline; although, the use of such a specific dataset - with not totally realistic scenes - is a limitation of our work.

\vspace{-4pt}
\section{Conclusion and Future Work}\label{sec:Conclusion}
\vspace{-4pt}


In this paper we propose a model for the sonification of silent videos by generating an audio track that is temporally and semantically aligned with the target video.
Our model is divided in two parts: a video onset network, with which onsets of actions present in an input silent video can be extracted; and a diffusion model that, conditioned on an onset track and a latent representation of the desired sound allows to generate a matching audio track that is synchronized to the onsets.

In future work, we plan to create a new dataset with audio-video pairs and onset annotations for scenes of interest in Foley generation. We'll extract these scenes from films and video games to test the model in realistic settings.
Additionally, we aim to explore novel approaches for training the onset model with minimal annotations, avoiding the need for manual annotation of every action in the video.
\vspace{-4pt}
\section{Acknowledgements}\label{sec:acknowledgement}
\vspace{-4pt}
M.C. is funded by UKRI and EPSRC as part of the ``UKRI CDT in Artificial Intelligence and Music'', under grant EP/S022694/1. The work of R. F. G. was partly supported by the PNRR MUR project ``Centro Nazionale 1 - Spoke 6'', under grant number CN1321845CE18353. E.P. is supported
by the ERC Grant no. 802554 (SPECGEO) and PRIN 2020 project no.2020TA3K9N (LEGO.AI).




\bibliographystyle{IEEEbib}
\bibliography{main}

\begin{thebibliography}{10}

\bibitem{Iashin2022SparseIS}
Vladimir~E. Iashin, Weidi Xie, Esa Rahtu, and Andrew Zisserman,
\newblock ``Sparse in space and time: Audio-visual synchronisation with
  trainable selectors,''
\newblock in {\em British Machine Vision Conference}, 2022.

\bibitem{Chen2021AudioVisualSI}
Honglie Chen, Weidi Xie, Triantafyllos Afouras, Arsha Nagrani, Andrea Vedaldi,
  and Andrew Zisserman,
\newblock ``Audio-visual synchronisation in the wild,''
\newblock {\em ArXiv}, vol. abs/2112.04432, 2021.

\bibitem{Kadandale2022VocaLiSTAA}
Venkatesh~Shenoy Kadandale, Juan~F. Montesinos, and Gloria Haro,
\newblock ``Vocalist: An audio-visual synchronisation model for lips and
  voices,''
\newblock {\em ArXiv}, vol. abs/2204.02090, 2022.

\bibitem{Dassani2019AutomatedCO}
Vansh Dassani, Jon Bird, and Dave Cliff,
\newblock ``Automated composition of picture-synched music soundtracks for
  movies,''
\newblock {\em Proceedings of the 16th ACM SIGGRAPH European Conference on
  Visual Media Production}, 2019.

\bibitem{Zhang2021RepetitiveAC}
Yunhua Zhang, Ling Shao, and Cees G.~M. Snoek,
\newblock ``Repetitive activity counting by sight and sound,''
\newblock {\em 2021 IEEE/CVF Conference on Computer Vision and Pattern
  Recognition (CVPR)}, pp. 14065--14074, 2021.

\bibitem{Yao2023PoseRACPS}
Ziyu Yao, Xuxin Cheng, and Yuexian Zou,
\newblock ``Poserac: Pose saliency transformer for repetitive action
  counting,''
\newblock {\em ArXiv}, vol. abs/2303.08450, 2023.

\bibitem{Dwibedi2020CountingOT}
Debidatta Dwibedi, Yusuf Aytar, Jonathan Tompson, Pierre Sermanet, and Andrew
  Zisserman,
\newblock ``Counting out time: Class agnostic video repetition counting in the
  wild,''
\newblock {\em 2020 IEEE/CVF Conference on Computer Vision and Pattern
  Recognition (CVPR)}, pp. 10384--10393, 2020.

\bibitem{Ghose2021FoleyGANVG}
Sanchita Ghose and John~J. Prevost,
\newblock ``Foleygan: Visually guided generative adversarial network-based
  synchronous sound generation in silent videos,''
\newblock {\em ArXiv}, vol. abs/2107.09262, 2021.

\bibitem{Su2023PhysicsDrivenDM}
Kun Su, Kaizhi Qian, Eli Shlizerman, Antonio Torralba, and Chuang Gan,
\newblock ``Physics-driven diffusion models for impact sound synthesis from
  videos,''
\newblock {\em 2023 IEEE/CVF Conference on Computer Vision and Pattern
  Recognition (CVPR)}, pp. 9749--9759, 2023.

\bibitem{Luo2023DiffFoleySV}
Simian Luo, Chuanhao Yan, Chenxu Hu, and Hang Zhao,
\newblock ``Diff-foley: Synchronized video-to-audio synthesis with latent
  diffusion models,''
\newblock {\em ArXiv}, vol. abs/2306.17203, 2023.

\bibitem{Wang2023V2AMapperAL}
Heng Wang, Jianbo Ma, Santiago Pascual, Richard Cartwright, and Weidong~(Tom)
  Cai,
\newblock ``V2a-mapper: A lightweight solution for vision-to-audio generation
  by connecting foundation models,''
\newblock {\em ArXiv}, vol. abs/2308.09300, 2023.

\bibitem{Du2023ConditionalGO}
Yuexi Du, Ziyang Chen, Justin Salamon, Bryan~C. Russell, and Andrew Owens,
\newblock ``Conditional generation of audio from video via foley analogies,''
\newblock {\em 2023 IEEE/CVF Conference on Computer Vision and Pattern
  Recognition (CVPR)}, pp. 2426--2436, 2023.

\bibitem{Mariani2023MultiSourceDM}
Giorgio Mariani, Irene Tallini, Emilian Postolache, Michele Mancusi, Luca~Di
  Cosmo, and Emanuele Rodol{\`a},
\newblock ``Multi-source diffusion models for simultaneous music generation and
  separation,''
\newblock {\em ArXiv}, vol. abs/2302.02257, 2023.

\bibitem{Pascual2022FullbandGA}
Santiago Pascual, Gautam Bhattacharya, Chunghsin Yeh, Jordi Pons, and Joan
  Serr{\`a},
\newblock ``Full-band general audio synthesis with score-based diffusion,''
\newblock {\em ArXiv}, vol. abs/2210.14661, 2022.

\bibitem{niizumi2021byol}
Daisuke Niizumi, Daiki Takeuchi, Yasunori Ohishi, Noboru Harada, and Kunio
  Kashino,
\newblock ``Byol for audio: Self-supervised learning for general-purpose audio
  representation,''
\newblock in {\em 2021 International Joint Conference on Neural Networks
  (IJCNN)}. IEEE, 2021, pp. 1--8.

\bibitem{saeed2021contrastive}
Aaqib Saeed, David Grangier, and Neil Zeghidour,
\newblock ``Contrastive learning of general-purpose audio representations,''
\newblock in {\em ICASSP 2021-2021 IEEE International Conference on Acoustics,
  Speech and Signal Processing (ICASSP)}. IEEE, 2021, pp. 3875--3879.

\bibitem{elizalde2023clap}
Benjamin Elizalde, Soham Deshmukh, Mahmoud Al~Ismail, and Huaming Wang,
\newblock ``Clap learning audio concepts from natural language supervision,''
\newblock in {\em ICASSP 2023-2023 IEEE International Conference on Acoustics,
  Speech and Signal Processing (ICASSP)}. IEEE, 2023, pp. 1--5.

\bibitem{song2021scorebased}
Yang Song, Jascha Sohl-Dickstein, Diederik~P Kingma, Abhishek Kumar, Stefano
  Ermon, and Ben Poole,
\newblock ``Score-based generative modeling through stochastic differential
  equations,''
\newblock in {\em International Conference on Learning Representations}, 2021.

\bibitem{salimans2022progressive}
Tim Salimans and Jonathan Ho,
\newblock ``Progressive distillation for fast sampling of diffusion models,''
\newblock in {\em International Conference on Learning Representations}, 2022.

\bibitem{song2021denoising}
Jiaming Song, Chenlin Meng, and Stefano Ermon,
\newblock ``Denoising diffusion implicit models,''
\newblock in {\em International Conference on Learning Representations}, 2021.

\bibitem{schneider2023mo}
Flavio Schneider, Zhijing Jin, and Bernhard Sch{\"o}lkopf,
\newblock ``Mo$\backslash$\^{} usai: Text-to-music generation with long-context
  latent diffusion,''
\newblock {\em arXiv preprint arXiv:2301.11757}, 2023.

\bibitem{ho2021classifierfree}
Jonathan Ho and Tim Salimans,
\newblock ``Classifier-free diffusion guidance,''
\newblock in {\em NeurIPS 2021 Workshop on Deep Generative Models and
  Downstream Applications}, 2021.

\bibitem{Owens2015VisuallyIS}
Andrew Owens, Phillip Isola, Josh~H. McDermott, Antonio Torralba, Edward~H.
  Adelson, and William~T. Freeman,
\newblock ``Visually indicated sounds,''
\newblock {\em 2016 IEEE Conference on Computer Vision and Pattern Recognition
  (CVPR)}, pp. 2405--2413, 2015.

\bibitem{kilgour2018fr}
Kevin Kilgour, Mauricio Zuluaga, Dominik Roblek, and Matthew Sharifi,
\newblock ``Fr$\backslash$'echet audio distance: A metric for evaluating music
  enhancement algorithms,''
\newblock {\em arXiv preprint arXiv:1812.08466}, 2018.

\bibitem{Perez2017TheEO}
Luis Perez and Jason Wang,
\newblock ``The effectiveness of data augmentation in image classification
  using deep learning,''
\newblock {\em ArXiv}, vol. abs/1712.04621, 2017.

\bibitem{SpecVQGAN_Iashin_2021}
Vladimir Iashin and Esa Rahtu,
\newblock ``Taming visually guided sound generation,''
\newblock in {\em British Machine Vision Conference (BMVC)}, 2021.

\bibitem{Kumar2019MelGANGA}
Kundan Kumar, Rithesh Kumar, Thibault de~Boissi{\`e}re, Lucas Gestin, Wei~Zhen
  Teoh, Jose M.~R. Sotelo, Alexandre de~Br{\'e}bisson, Yoshua Bengio, and
  Aaron~C. Courville,
\newblock ``Melgan: Generative adversarial networks for conditional waveform
  synthesis,''
\newblock in {\em Neural Information Processing Systems}, 2019.

\end{thebibliography}

\end{document}